\documentclass[angeo]{copernicus}

\usepackage{color}
\usepackage{url}

\newcommand{\zeroindexSmall}{{\rm o}}

\newcommand{\nzero}{n_\zeroindexSmall}
\newcommand{\vzero}{v_\zeroindexSmall}
\newcommand{\epsilonzero}{\epsilon_\zeroindexSmall}

\newcommand{\MARKII}[1]{#1}

\begin{document}


\title{Simulation study of the plasma brake effect}

\author{P.~Janhunen}

\affil{Finnish Meteorological Institute, POB-503, FI-00101, Helsinki, Finland}


\runningtitle{Plasma brake PIC simulation}

\runningauthor{P.~Janhunen}

\correspondence{Pekka Janhunen\\ (pekka.janhunen@fmi.fi)}

\received{}
\pubdiscuss{} 
\revised{}
\accepted{}
\published{}


\firstpage{1}

\maketitle  

\begin{abstract}
The plasma brake is a thin negatively biased tether which has been
proposed as an efficient concept for deorbiting satellites and debris
objects from low Earth orbit. We simulate the interaction with the
ionospheric plasma ram flow with the plasma brake tether by a high
performance electrostatic particle in cell code to evaluate the
thrust. The tether is assumed to be perpendicular to the flow. We
perform runs for different tether voltage, magnetic field orientation
and plasma ion mass. We show that a simple analytical thrust formula
reproduces most of the simulation results well. The interaction with
the tether and the plasma flow is laminar \MARKII{(i.e., smooth and not turbulent)} when the magnetic field is
perpendicular to the tether and the flow. If the magnetic field is
parallel to the tether, the behaviour is unstable and thrust is
reduced by a modest factor. The case when the magnetic field is
aligned with the flow can also be unstable, but does not result in
notable thrust reduction. We also fix an error in an earlier
reference. According to the simulations\MARKII{,} the predicted thrust of the
plasma brake is large enough to make the method promising for
\MARKII{low Earth orbit (LEO)} satellite deorbiting. As a numerical example we estimate that a 5 km
long plasma brake tether weighing 0.055 kg could produce 0.43 mN
breaking force which is enough to reduce the orbital altitude of a 260
kg \MARKII{object} mass by 100 km during one year.
\end{abstract}


\introduction  

The plasma brake \citep{paper3,Plasmabrake} is an efficient
propellantless concept for deorbiting low Earth orbit (LEO)
satellites. The plasma brake is a very thin negatively charged tether
which, when charged, causes a braking force by creating enhanced
Coulomb drag with ambient ionospheric plasma ram flow. The plasma
brake tether is somewhat similar to the more well known electrodynamic
tether \citep{SanmartinEtAl1993}, but is much thinner and uses electrostatic rather than
magnetic forces.

Because the plasma brake 4-wire ultrasonically bonded tether
is thin, it is lightweight, 11 grams per kilometre
\citep{SeppanenEtAl2013} and does not form an appreciable threat to
other satellites even in case of tether breakage. A broken piece of
plasma brake tether deorbits itself quickly because of electromagnetic
forces and neutral drag, and even if the tether piece would collide
with a satellite, the 25-50 $\mu$m wires draw only $\sim 0.1$ mm deep
scratches on its surface. \MARKII{Said electromagnetic forces
  are a passive Coulomb drag effect and to a lesser extent
  the passive electrodynamic tether effect. The broken tether piece
  experiences an orbital motion induced natural electric field, and
  whenever this field has a component along the tether, one end of the
tether gets charged negatively and the other one positively with respect to
the local plasma, causing electrostatic Coulomb drag and
electromagnetic Lorentz drag. Although the passive electromagnetic drag effects are weaker than the active
ones, the very low specific mass of the tether ($\sim 0.01$ kg/km)
makes passive deorbiting of a broken tether piece fast, typically.}

\MARKII{To maintain a tether at positive voltage in the solar wind
  requires an
  electron gun \citep{RSIpaper,paper7a} that pumps out negatively charged
  particles from the system and thereby cancels the tether's gathered plasma current.
  Likewise, to maintain a tether at negative
  voltage can be accomplished with a positive ion gun. However, in
  LEO the maintenance of} the plasma brake
tether's negative voltage in most cases
does not require an ion gun since the satellite's conducting body can
be used as the current balancing electron gathering surface
\citep{Plasmabrake}. If the satellite's grounded and conducting surface
area is insufficiently small for this purpose, a relatively short
positively biased tether made of similar material than the main
tether could be used for gathering the required balancing electron
current.

The topic of the paper is to use a realistic electrostatic particle in
cell (PIC) simulation to model the interaction between the negatively
biased plasma brake tether with the surrounding ionospheric plasma ram
flow which moves at a typical satellite orbital speed of 7.5 km/s. We
run the code with different voltages, different magnetic field
orientations and different ion species and determine the thrust per
tether length in each case. We will also present a simple analytical
formula for plasma brake thrust which reproduces the simulation
results.

\section{Simulation code}

We use a two-dimensional explicit electrostatic kinetic PIC simulation
model \citep{BirdsallAndLangdon}. The functionality of the code is
similar to what we have used earlier to model the
positively biased E-sail tether \citep{paper2}, but the code is
vectorised with AgnerVector library \citep{AgnerVector} and
parallelised with the standard Message Passing Interface (MPI) and
OpenMP tools. We run the code with real electron and ion masses.

We simulate a negatively charged tether in plasma flow which is
\MARKII{caused by} the satellite's orbital motion through the ionosphere. We use a
Cartesian coordinate system where $X$ is antiparallel with the
\MARKII{tether-perpendicular component of the} flow, $Z$ is parallel
to the tether and $Y$ completes a right-handed coordinate system.
\MARKII{For convenience we take the plasma flow to be perpendicular to the tether,
  which implies no loss in generality.} The
simulation is two-dimensional in the tether-perpendicular plane. A
constant external magnetic field is employed in some runs. By default
the plasma is cold oxygen plasma, temperature 0.1 eV, and it flows at
7.5 km/s which corresponds to 4.7 eV bulk flow ion energy. The tether
bias voltage is typically a few hundred volts negative. The parameters
of all reported runs are given in Table \ref{tab:SimParams}. The
simulations are initialised from vacuum and the plasma flow starts
entering the box at $t=0$.

\MARKII{Above $\sim$ 500 km where the plasma brake is relevant
(at lower altitude the neutral drag typically brings satellites down
rapidly enough), ionospheric plasma consists mainly of O$^{+}$ (16 \MARKII{amu}), N$^{+}$
(14 \MARKII{amu}), He$^{+}$ (4 \MARKII{amu}) and H$^{+}$ (1 \MARKII{amu}) ions. The main species are O$^{+}$ and H$^{+}$;
the minor species N$^{+}$ can be summed to O$^{+}$ because the masses are
approximately similar. The abundance of O$^{+}$ decreases with altitude. At
all altitudes, solar activity tends to increase the oxygen
abundance. For the plasma brake, the most relevant environment is O$^{+}$
plasma because proton plasma produces order of magnitude smaller
thrust and helium is usually not a dominant species. For this reason,
the main emphasis in this paper is on O$^{+}$ plasma.}

Runs reported in this paper used 16 nodes on a Cray XC30. The XC30
consists of compute nodes, each node has 2 processor chips and each
chip has 10 2.6 GHz execution cores. The runs reported here used 16
XC30 nodes and achieved $4.4\cdot 10^9$ particle propagations per
second which corresponds to $\sim 1$ Tflops single precision floating
point performance.

\section{Results}

\subsection{Oxygen plasma without magnetic field}

We first run the code with oxygen plasma and without magnetic field
for different values of the tether potential $V_w$. The runs are
detailed in Table \ref{tab:RunList}. The vacuum potential of the
tether \MARKII{(the voltage difference between the tether and the
  simulation box boundary if there is no plasma inside the box)} is an input parameter to the simulation. After the simulation,
the true potential of the tether is calculated by evaluating the
plasma potential at the tether and adding it to the vacuum
potential. For this reason\MARKII{,} the tether potential values $V_w$ are not
round numbers. We label the runs partly by their vacuum potential values.

\begin{figure}[t]
\vspace*{2mm}
\begin{center}
\includegraphics[width=8.3cm]{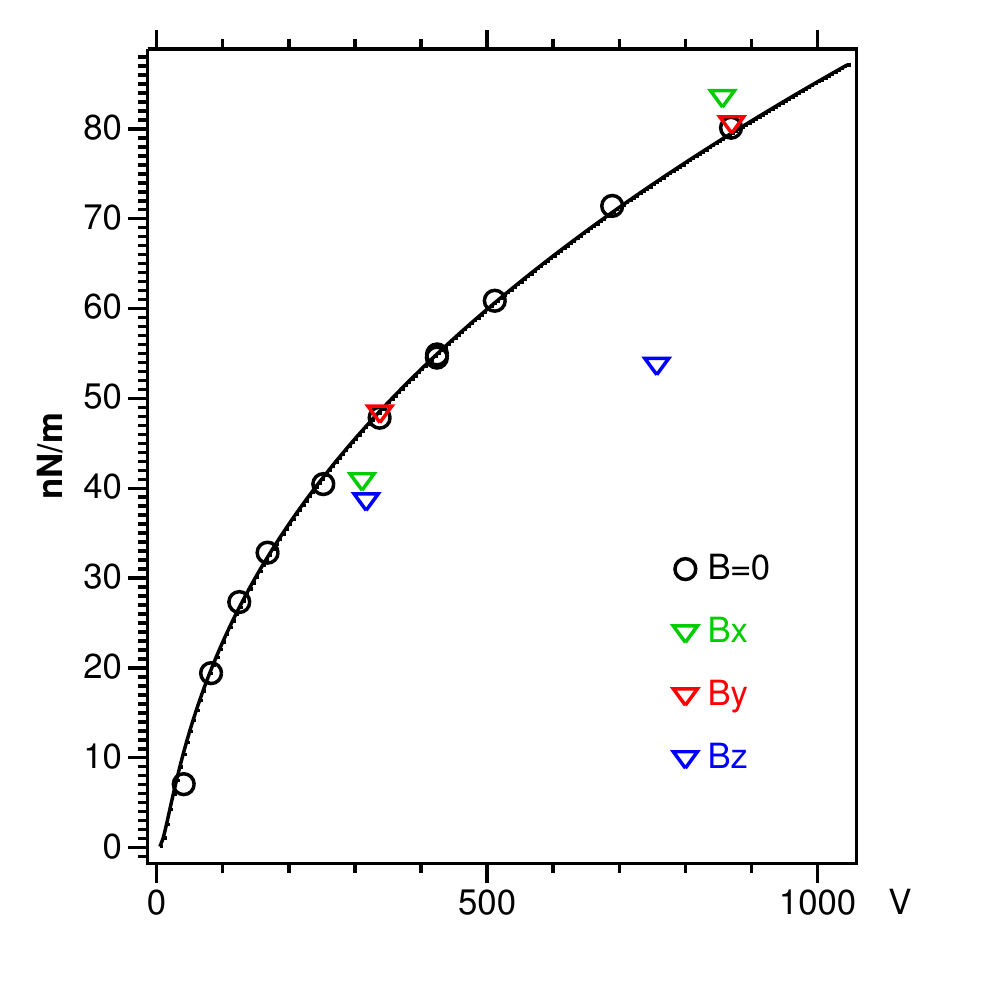}
\end{center}
\caption{Different marks: thrust as function of tether voltage $V_w$ in
  oxygen plasma with and without magnetic field, see legend in plot
  and Table \ref{tab:RunList} for run parameters. Solid line: equation
  (\ref{eq:dFdz}).
}
\label{fig:Vdep}
\end{figure}

Figure \ref{fig:Vdep} shows the simulated thrust (determined from
particle momentum balance averaged over the last 1/3 of the run) with
open circles. Coloured marks in Fig.~\ref{fig:Vdep} are results from
some runs containing a magnetic field and they are described in later
subsections below. The solid line is the following analytical formula,
inspired by our earlier work \citep[equation 20]{paper3}:
\begin{equation}
\frac{dF}{dz} =
3.864 \times P_{\rm dyn} \sqrt{\frac{\epsilonzero \tilde{V}}{e \nzero}}
\exp\left(-V_i/\tilde{V}\right)
\label{eq:dFdz}
\end{equation}
where $P_{\rm dyn} = m_i \nzero \vzero^2$ is the dynamic pressure,
$m_i$ is the ion mass ($m_i=16$ \MARKII{amu} for oxygen plasma here), $\vzero$ is
the plasma flow speed relative to spacecraft (assumed to be perpendicular to
the tether or else $\vzero$ denotes only the perpendicular component),
\begin{equation}
\tilde{V} = \frac{V_w}{\ln(\lambda_D^{\rm eff}/r_w^{*})},
\end{equation}
$r_w^{*}$ is the tether's effective electric radius \citep[appendix
  A]{paper2}, $\lambda_D^{\rm eff}=\sqrt{\epsilonzero V_w/(e \nzero)}$
is the effective Debye length and $V_i=(1/2)m_i \vzero^2/e$ is the
bulk ion flow energy in voltage units. The effective electric radius
is approximately given by $r_w^{*} = \sqrt{b r_w}$ where $r_w$ is the
tether wire radius, typically 12.5-25 $\mu$m, and $b$ is the tether
width, typically 2 cm (a rough value of $b$ is sufficient to
  know because $r_w^*$ enters into
  Eq.~(\ref{eq:dFdz}) only logarithmically). In this paper we use the value $r_w^{*}=1$
mm. The numerical coefficient (3.864) in front of the expression has
been selected to give a good fit to the present simulation
results. This value is about 2.25 times larger than the value 1.72
used in our previous work \citep{paper3}. We think that this
difference \MARKII{may} arise because in the earlier work we used typical solar
wind parameters so that the ratio $V_w/V_i$ was about 10 while this
ratio is about 100 in the present ionospheric case \MARKII{so that the
  works explored different regions of the parameter space}. In the ionospheric
case, many ions are deflected backward by the potential well \MARKII{around} the
tether which increases the thrust markedly in comparison to the solar
wind case where the much larger bulk flow speed causes the ions to
deflect more modestly.

\begin{figure}[t]
\vspace*{2mm}
\begin{center}
\includegraphics[width=8.3cm]{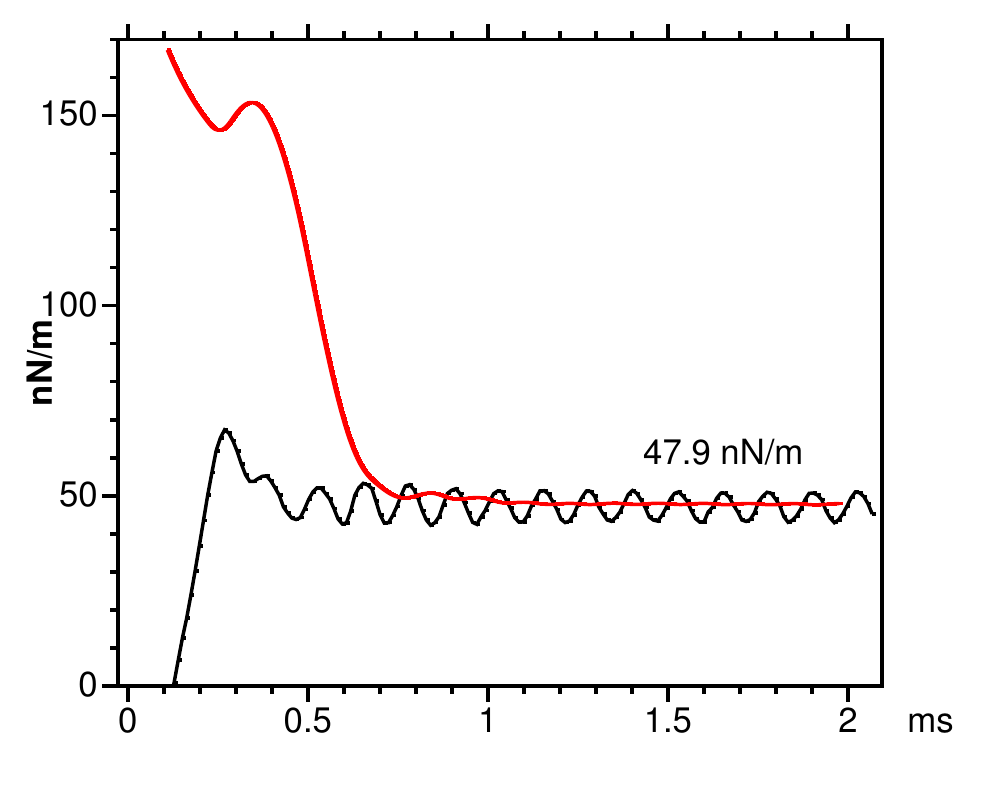}
\end{center}
\caption{Time development of thrust in V400 (Baseline) run, tether
  Coulomb force computation (black curve) and momentum balance
  computation (red curve).
}
\label{fig:CurveBaseline}
\end{figure}

\begin{figure}[t]
\vspace*{2mm}
\begin{center}
\includegraphics[width=8.3cm]{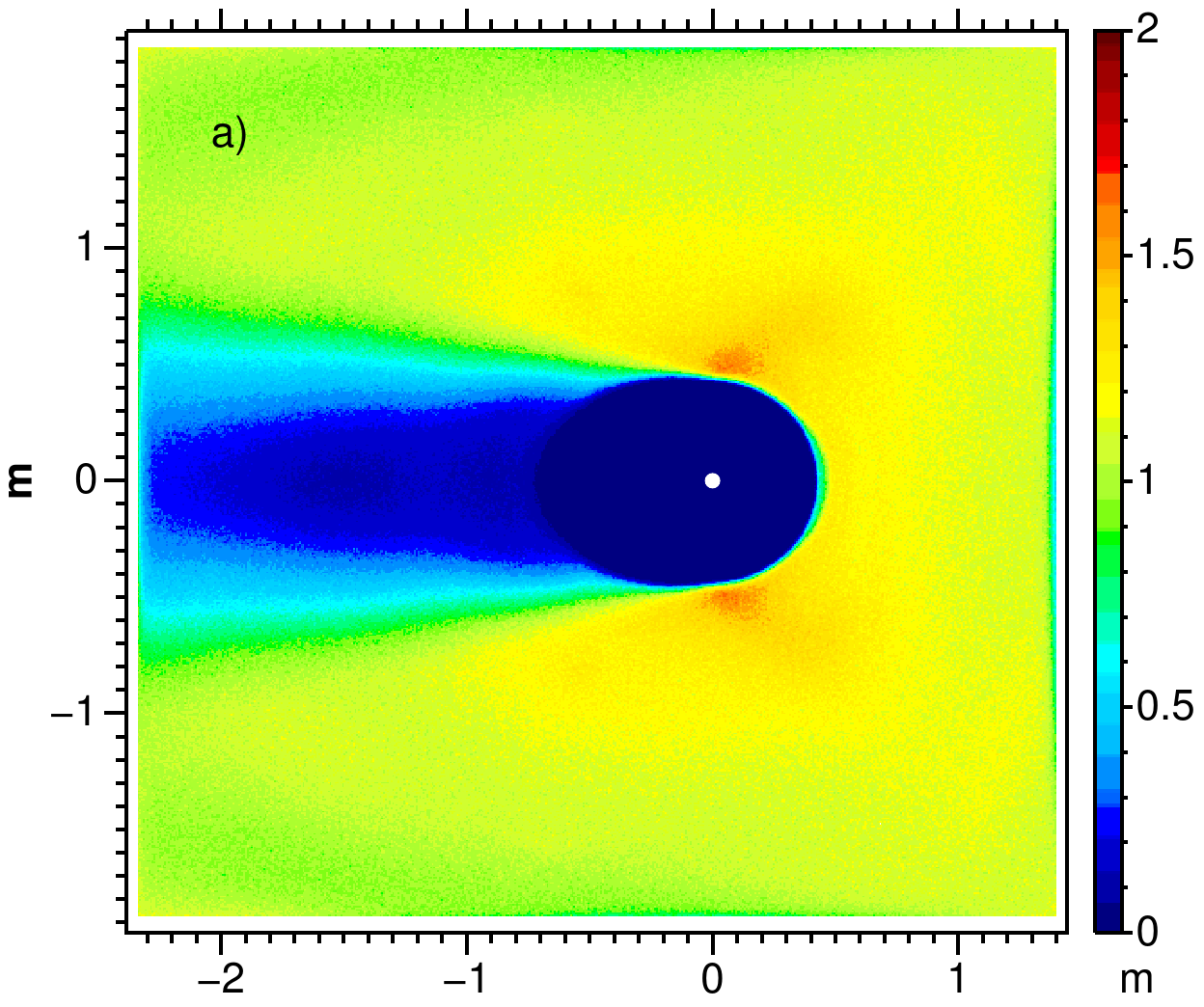}
\includegraphics[width=8.3cm]{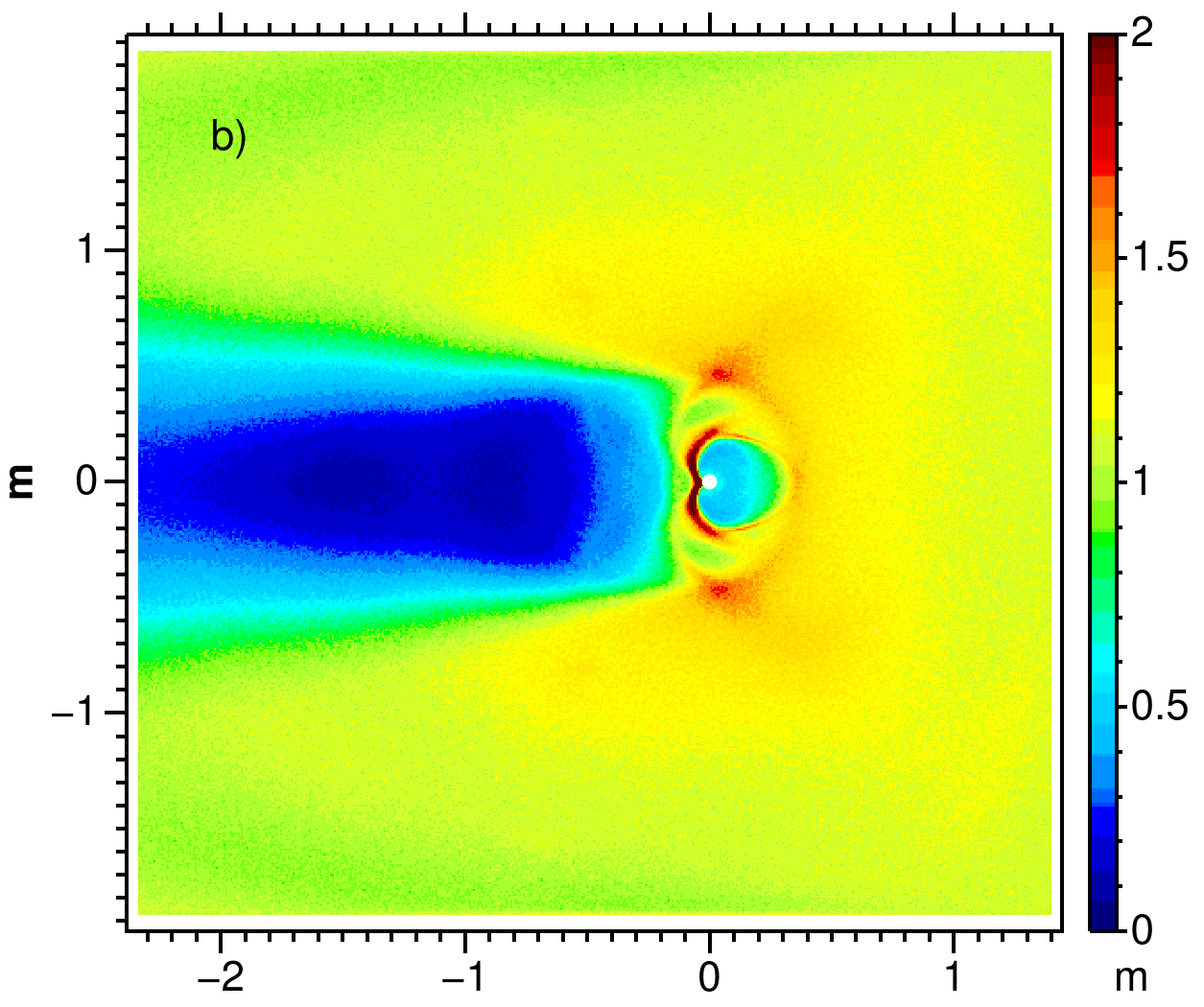}
\end{center}
\caption{Electron density (a) and ion density (b) normalised to plasma
  stream density in the final state ($t=2.43$ ms) of run V400 (Baseline).}
\label{fig:BaselinePcolor}
\end{figure}

Figure \ref{fig:CurveBaseline} shows the time history of thrust during
the V400 (Baseline) run, computed by two complementary methods: the
direct Coulomb force acting on the tether (by evaluating the numerical
gradient of the plasma potential at the tether position) and the total
momentum $X$ component lost by particles between entering and leaving
the simulation box. Figure \ref{fig:BaselinePcolor} shows the
two-dimensional instantaneous electron and ion density at the final
state of the Baseline run.

Increasing the tether voltage makes the ion sheath larger. The highest
voltage runs (marked with ``g'' in Table \ref{tab:RunList}) were
performed with extended box size of 768$\times$768 to accommodate the
larger sheath. As a consistency check, run V500 was performed with
both grid sizes. The difference in the determined thrust was minimal
(Table \ref{tab:RunList}, compare V500 and V500g).

The quasi-monochromatic oscillation seen in Fig.~\ref{fig:CurveBaseline}
occurs in runs V300 and higher. In lower voltage runs the oscillation
is absent. When present, the oscillation neither increases nor
decreases with time.

\subsection{Effect of magnetic field}

\begin{figure}[t]
\vspace*{2mm}
\begin{center}
\includegraphics[width=8.3cm]{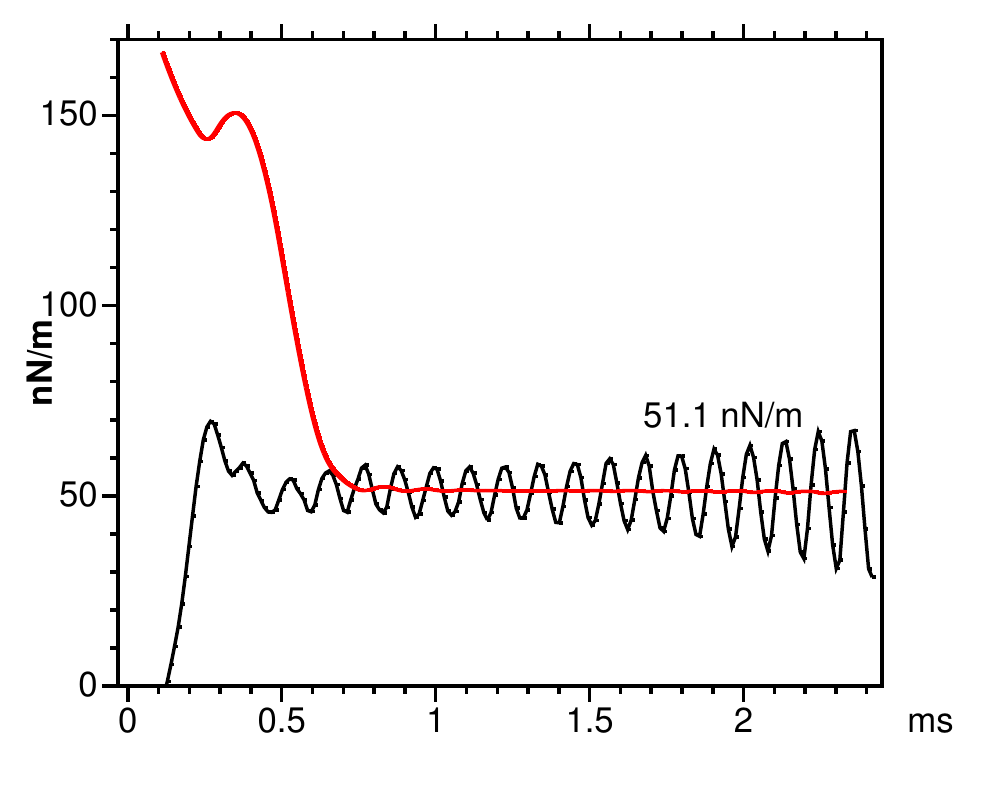}
\end{center}
\caption{Run Bx: same as Fig.~\ref{fig:CurveBaseline}, but with $B_x=30000$ nT.}
\label{fig:CurveBx}
\end{figure}

\begin{figure}[t]
\vspace*{2mm}
\begin{center}
\includegraphics[width=8.3cm]{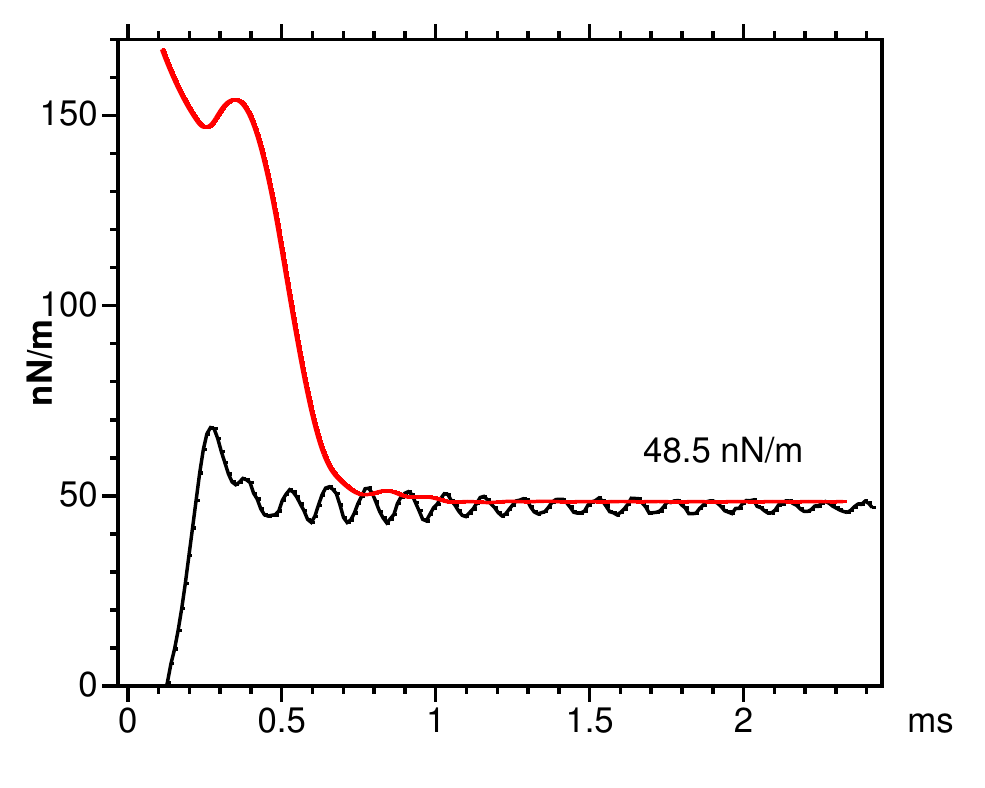}
\end{center}
\caption{Run By: same as Fig.~\ref{fig:CurveBaseline}, but with $B_y=30000$ nT.}
\label{fig:CurveBy}
\end{figure}

\begin{figure}[t]
\vspace*{2mm}
\begin{center}
\includegraphics[width=8.3cm]{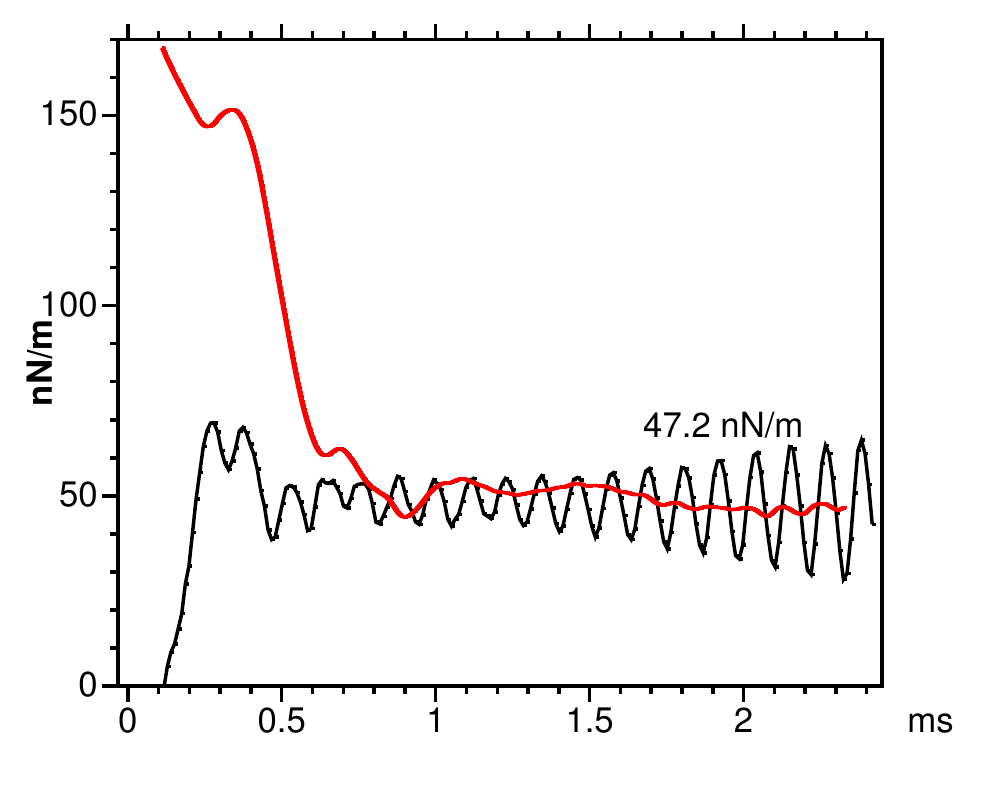}
\end{center}
\caption{Run Bz: same as Fig.~\ref{fig:CurveBaseline}, but with $B_z=30000$ nT.}
\label{fig:CurveBz}
\end{figure}

We performed three more runs with the same parameters as V400
(Baseline), but with each of the magnetic field components in turn set
to 30000 nT, a typical LEO field strength. The results are shown in
Figs.~\ref{fig:CurveBx}-\ref{fig:CurveBz}. It is seen that in runs Bx
and Bz, the oscillation which was already present in the Baseline run
is now increasing in amplitude. In contrast, in the By run the
oscillation is damped and is going to disappear. Thus, $B_y$ (magnetic
field perpendicular to tether and flow) is stabilising while
flow-directed or tether-directed field is destabilising at the
baseline voltage.

\begin{figure}[t]
\vspace*{2mm}
\begin{center}
\includegraphics[width=8.3cm]{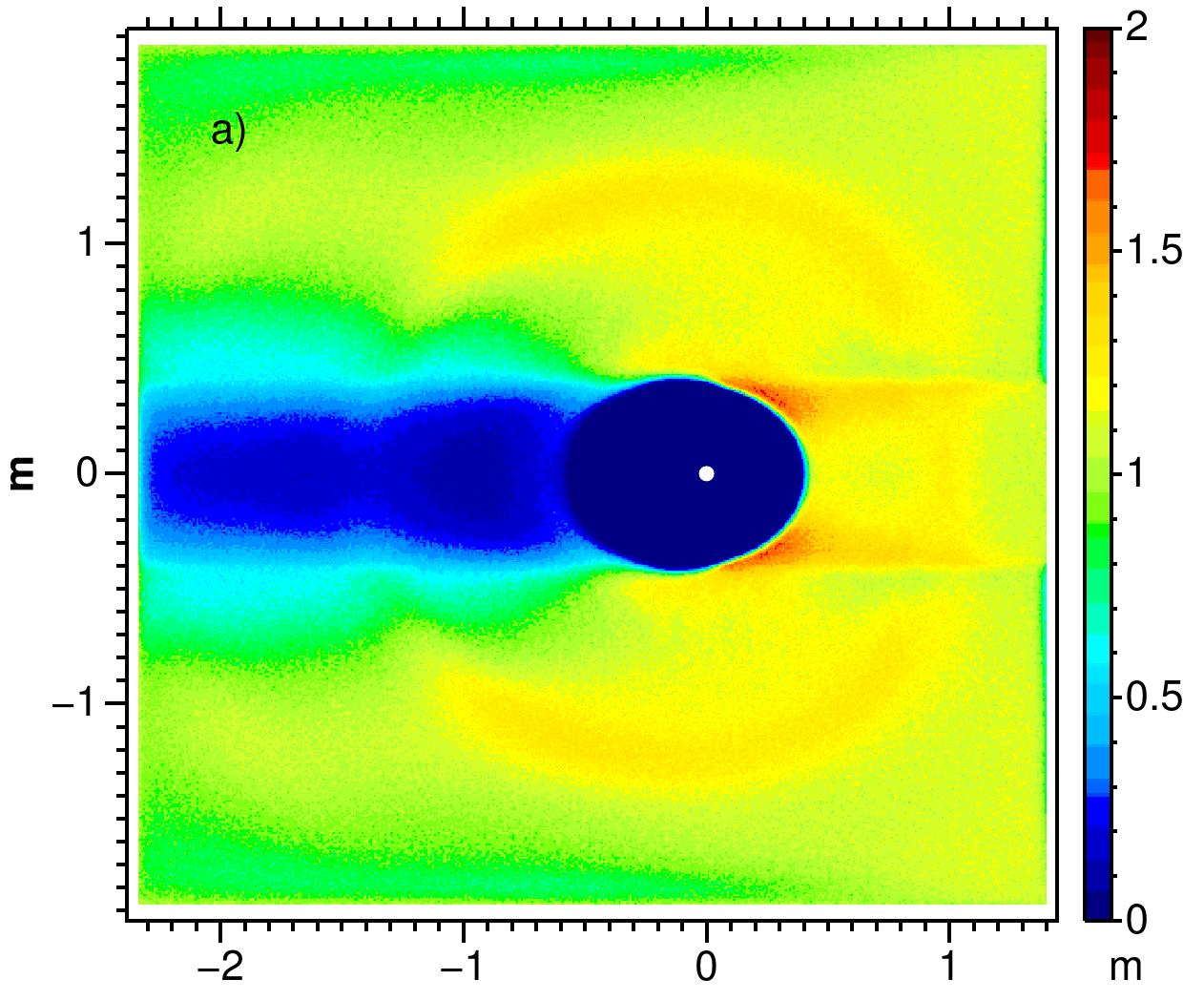}
\includegraphics[width=8.3cm]{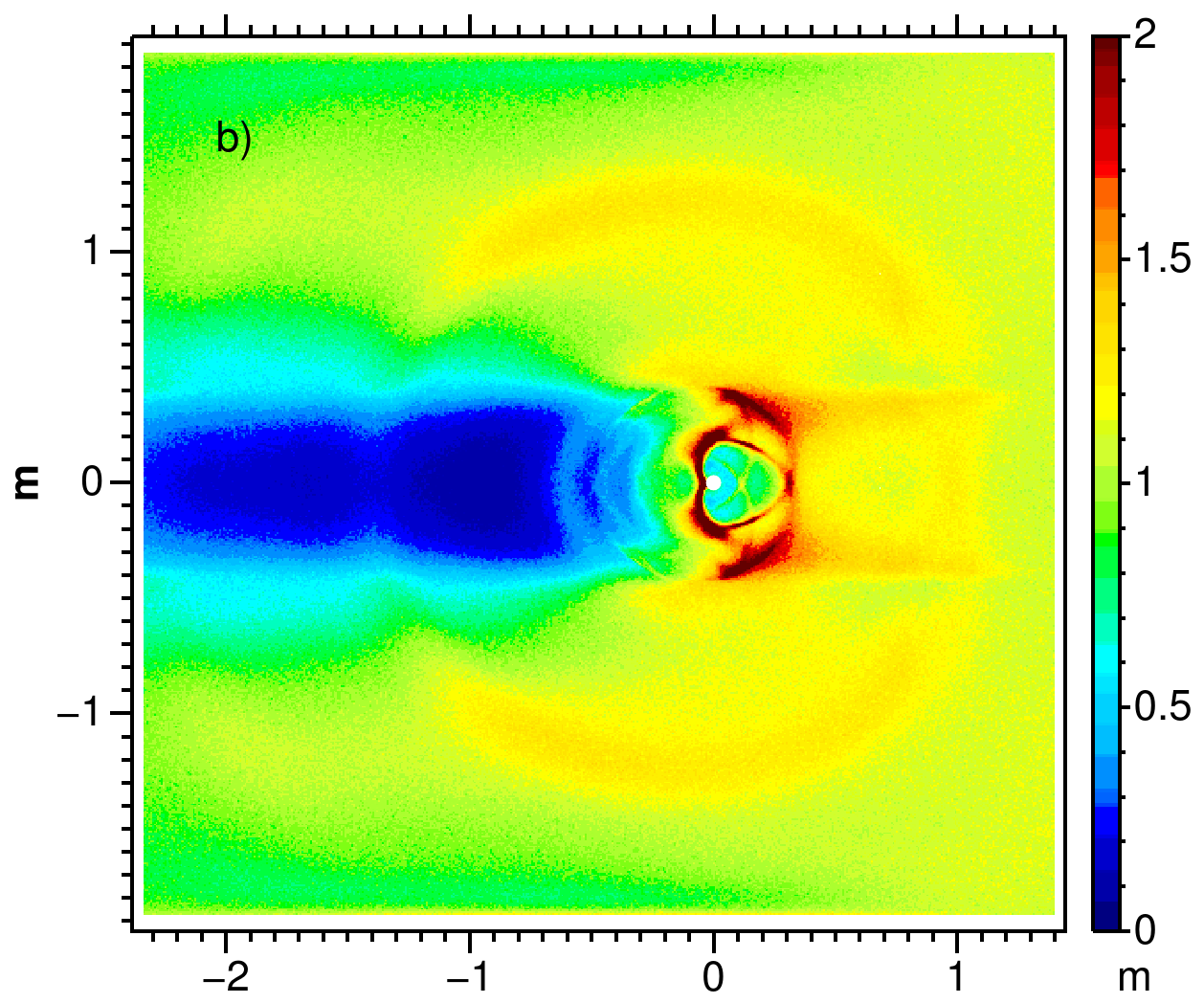}
\end{center}
\caption{Electron density (a) and ion density (b) normalised to plasma
  stream density in the final state ($t=2.43$ ms) of run Bx.}
\label{fig:BxPcolor}
\end{figure}

The final state of run Bx is shown in Fig.~\ref{fig:BxPcolor}. The
sheath surrounding the tether is unstable and radiates plasma waves in
all directions \MARKII{in the $XY$ plane}. In the tether-fixed coordinate frame, the waves move
approximately at the same phase speed as the bulk flow (7.5 km/s). The
waves \MARKII{appear} to move with the ion flow which is reflected by the
tether's potential well. The incoming flow has a Mach number of
$\approx 5$ with respect to ion acoustic wave speed. Because of energy
conservation, reflected ions move radially outward with the same speed
as the incoming flow. The tether's potential well fluctuates, which
modulates the flow of outward reflected ions. The modulations
propagate outward approximately at the same speed as the ions because
the ion acoustic speed is less than the bulk velocity of the outward
moving ion population. In Fig.~\ref{fig:BxPcolor} one can also see
that the $X$-directed magnetic field tends to restrict
electron motion \MARKII{in $Y$,} which creates some horizontal stripes
\MARKII{seen in the figure} (reddish stripes
on positive $X$), visible in both electron and ion density and
emanating from the edges of the electron cavity.

\begin{figure}[t]
\vspace*{2mm}
\begin{center}
\includegraphics[width=8.3cm]{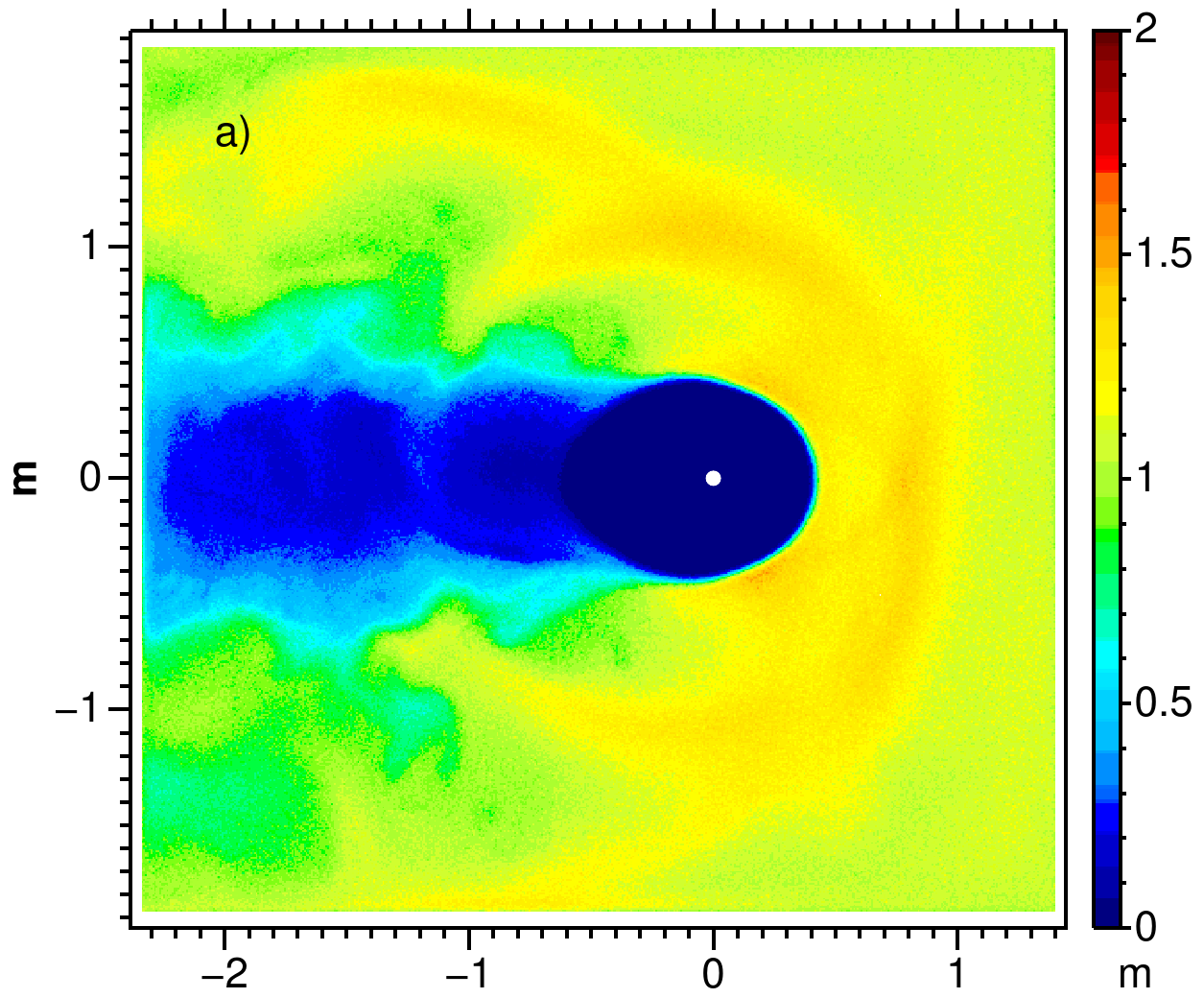}
\includegraphics[width=8.3cm]{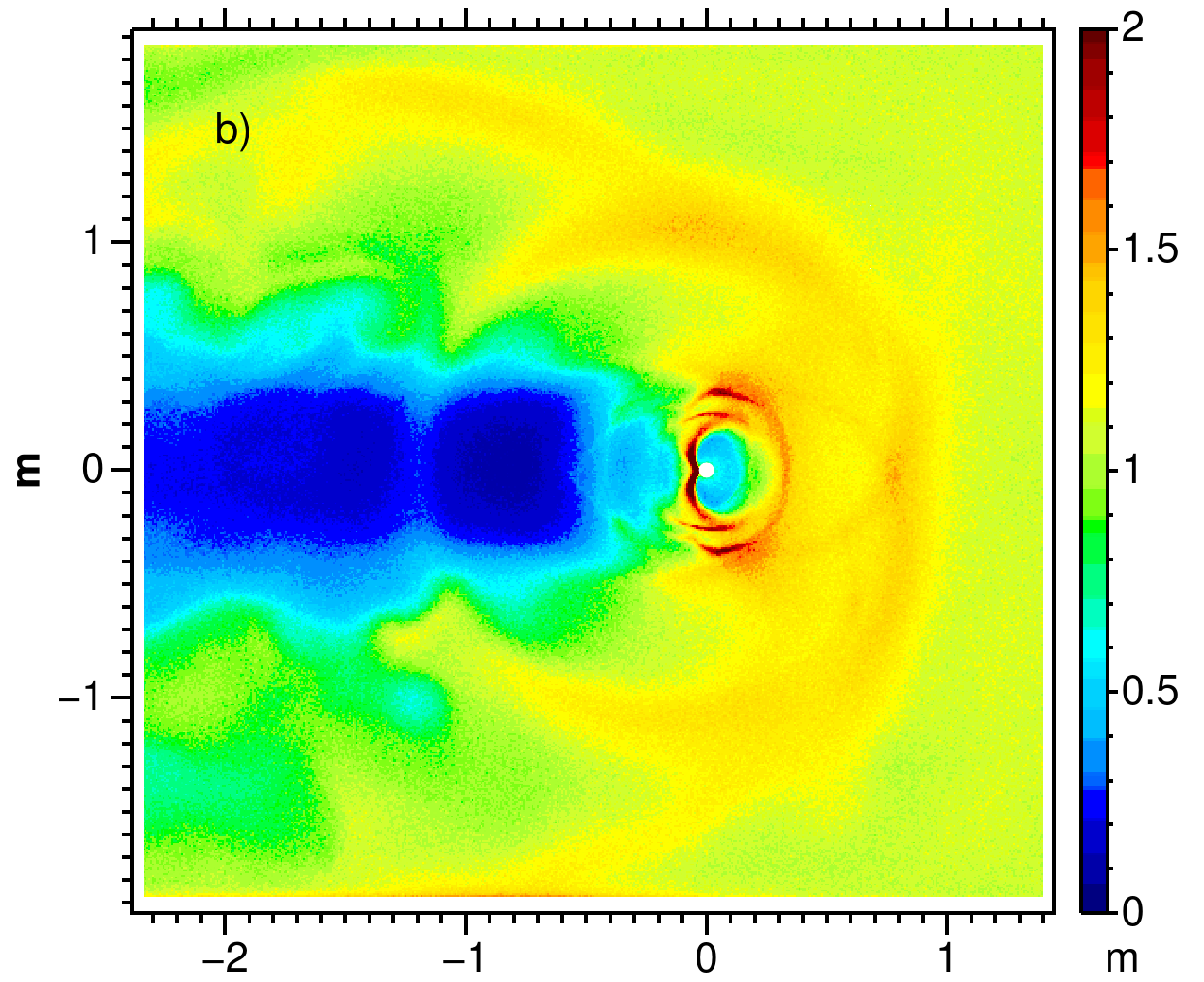}
\end{center}
\caption{Electron density (a) and ion density (b) normalised to plasma
  stream density in the final state ($t=2.43$ ms) of run Bz.}
\label{fig:BzPcolor}
\end{figure}

Figure \ref{fig:BzPcolor} shows the final state of run Bz. Again the
state is unstable and the tether's potential well radiates plasma
waves. Additionally, the boundaries of the plasma wake behind the
tether display some turbulent behaviour. This probably occurs because
now the magnetic field is along the tether so that electrons respond
to $X$ and $Y$ directed electric fields by ${\bf E}\times{\bf B}$ drifting
(similar to Kelvin-Helmholtz instability in plane perpendicular to
${\bf B}$). Not surprisingly, the horizontal stripes which were
visible in the Bx run (Fig.~\ref{fig:BxPcolor}) are now absent.

\subsection{Effect of ion mass}
\label{subsection:ionmass}

\begin{figure}[t]
\vspace*{2mm}
\begin{center}
\includegraphics[width=8.3cm]{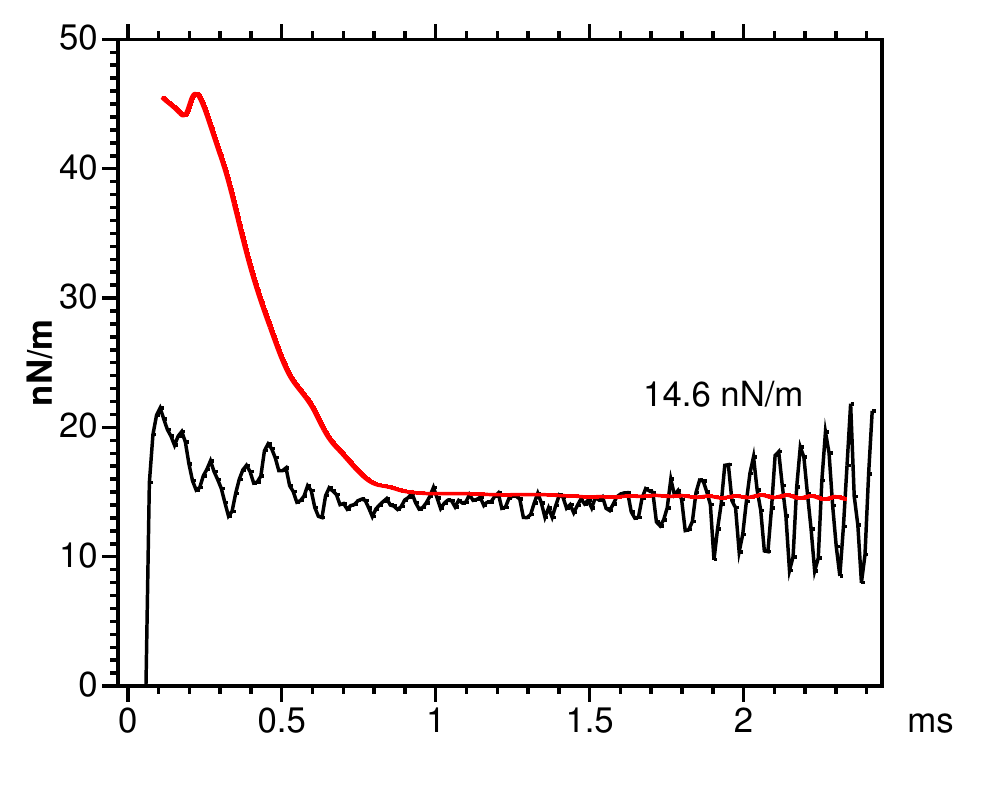}
\end{center}
\caption{Run Helium: same as Fig.~\ref{fig:CurveBaseline}, but in helium plasma.}
\label{fig:CurveHelium}
\end{figure}

\begin{figure}[t]
\vspace*{2mm}
\begin{center}
\includegraphics[width=8.3cm]{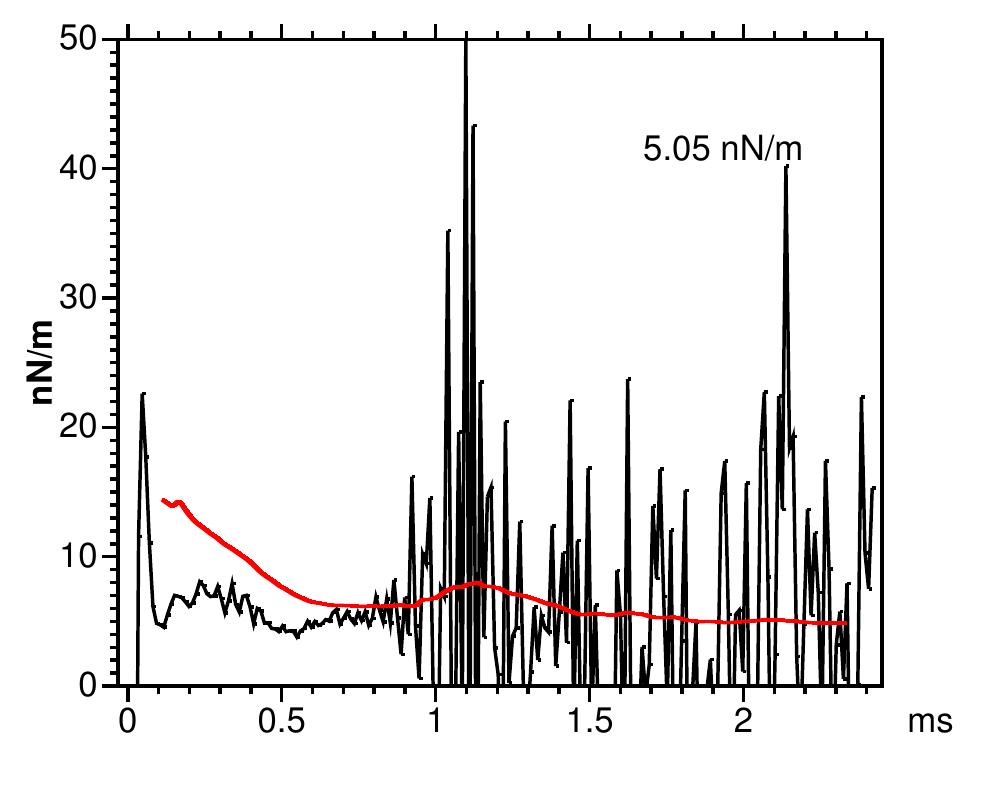}
\end{center}
\caption{Run Proton: same as Fig.~\ref{fig:CurveBaseline}, but in proton plasma.}
\label{fig:CurveProton}
\end{figure}

Figures \ref{fig:CurveHelium} and \ref{fig:CurveProton} show
thrust histories for helium and proton plasma, respectively, and with
other parameters in their Baseline (V400) values. In both light ion
cases, an unstable oscillation is present. The instability grows
faster in the proton run than in the helium run and has time to evolve
into intermittent, nonlinear regime.

\begin{figure}[t]
\vspace*{2mm}
\begin{center}
\includegraphics[width=8.3cm]{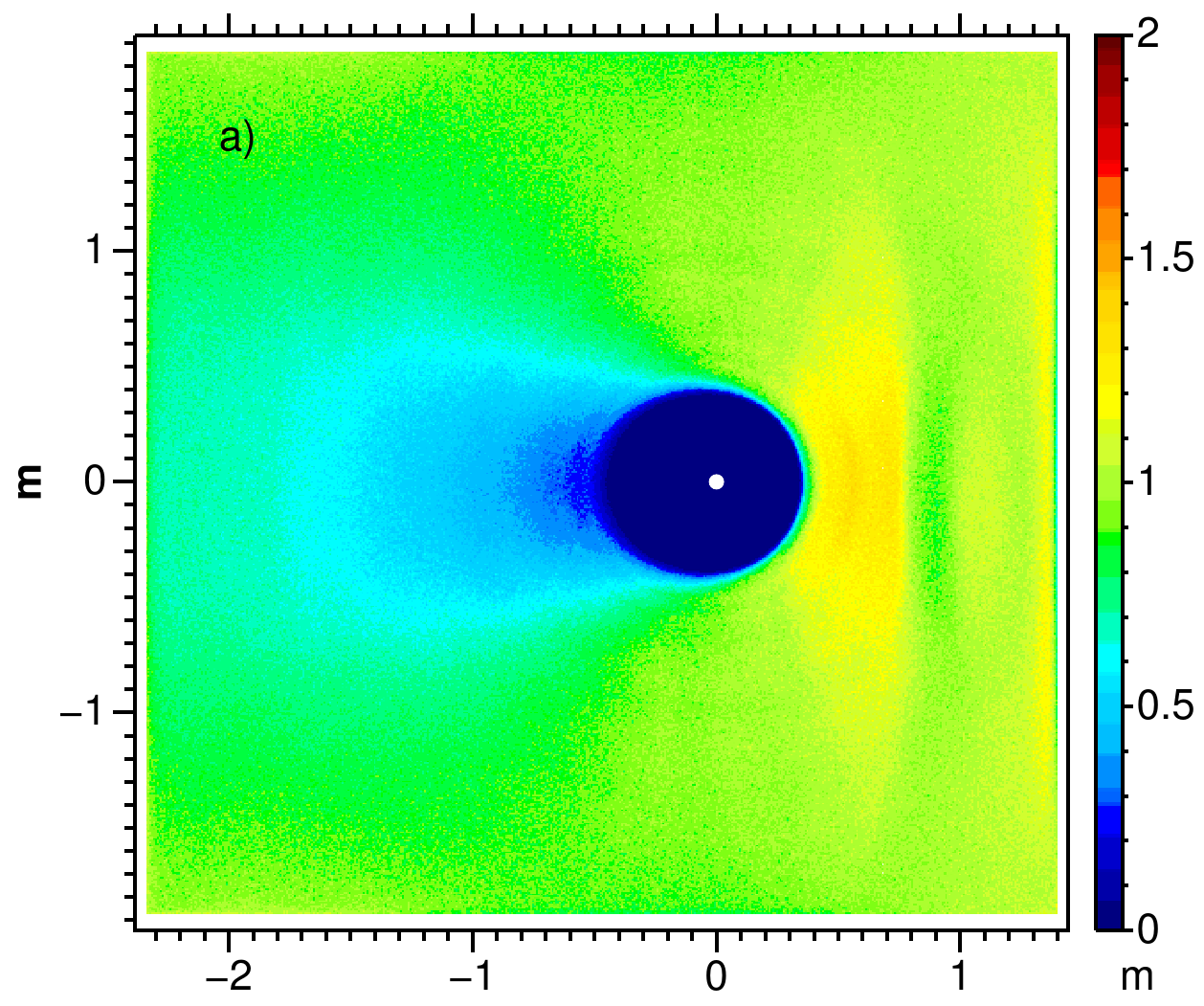}
\includegraphics[width=8.3cm]{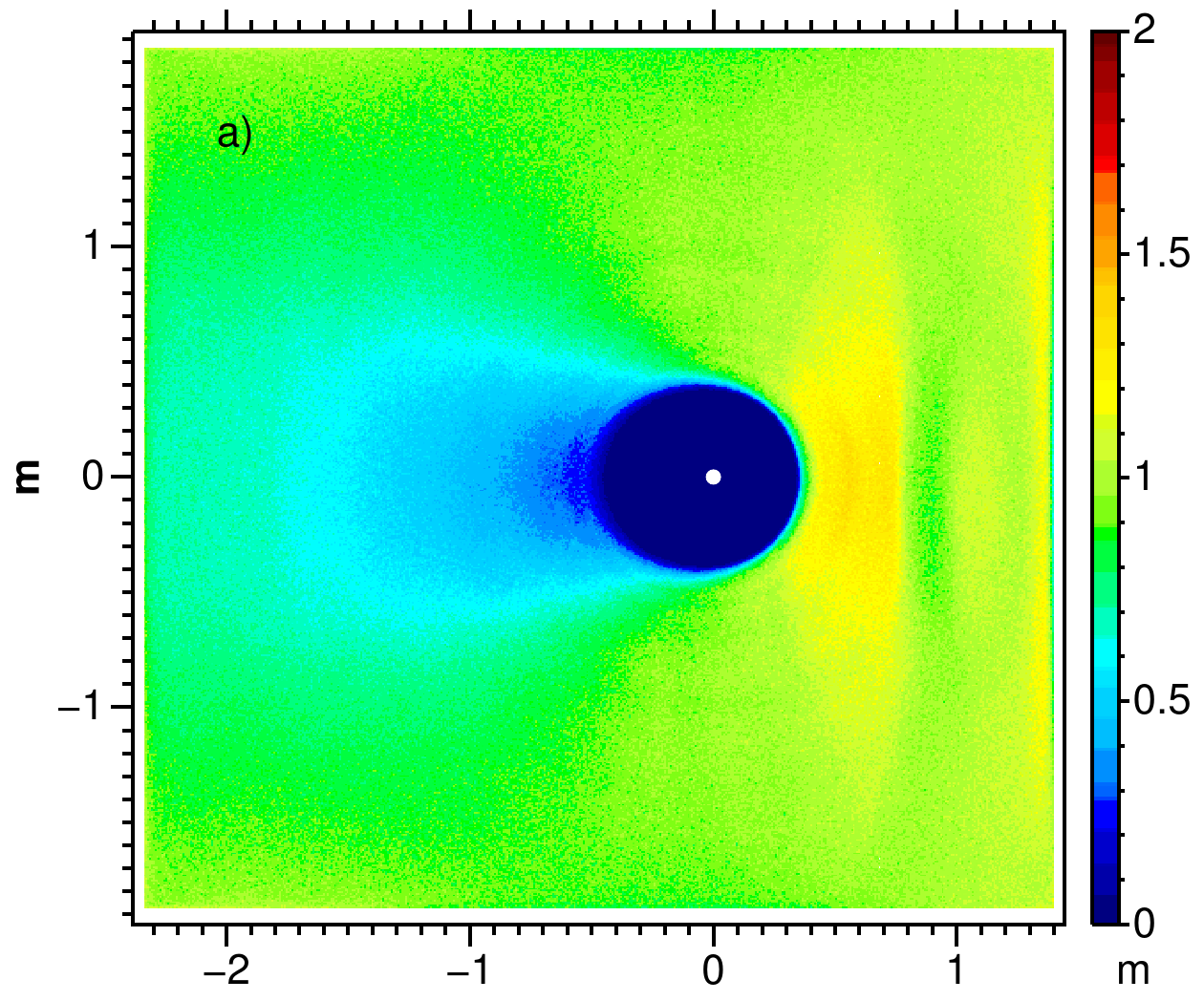}
\end{center}
\caption{Electron density (a) and ion density (b) normalised to plasma
  stream density in the final state ($t=2.43$ ms) of run Proton.}
\label{fig:ProtonPcolor}
\end{figure}

Figure \ref{fig:ProtonPcolor} shows the final state of the Proton
run. Some ions have become trapped by the potential well which has
decreased the thrust to some extent (Fig.~\ref{fig:CurveProton}). The
sheath emits plasma waves in all \MARKII{$XY$} directions. The flow Mach number with
respect to ion acoustic speed is \MARKII{low} (1.2) which is probably the
reason \MARKII{why} the wake behind the tether is rather short
\MARKII{(the wake gets filled quickly from the boundaries because the ion thermal speed is
  almost as large as the flow speed)}. In
Fig.~\ref{fig:ProtonPcolor} the behaviour of the densities near the
inflow boundary (right side) \MARKII{shows some $Y$-directed striping and} does not look completely natural. It is
possible that the finite box size affects the result to some extent
in this case. In the proton case (Fig.~\ref{fig:ProtonPcolor}) it is
also noteworthy that the emitted ion waves have short enough
wavelength that the electron density does not follow the ion density,
in other words, that the emitted wavelength is not much larger than
the electron Debye length.

Formula \ref{eq:dFdz} predicts 12.2 nN/m in the helium case and 2.92
nN/m in the proton case. The simulated values at 2.43 ms are larger
(14.6 and 5.05 nN/m, respectively), even though in the proton run some
ion trapping had already occurred which had reduced the
thrust. Probably the smaller dynamic pressure of light ion bulk flow
depresses the sheath less than in the oxygen case, and this effect is
not included in Eq.~(\ref{eq:dFdz}) where the thrust is assumed to be
linearly proportional to the ion mass. In the helium run (not shown)
the behaviour is intermediate between the oxygen and proton runs. In
subsection \ref{subsection:longerruns} below we will investigate
longer time behaviour of proton plasma where also a magnetic field is
included.

\subsection{Effect of electron temperature}

\begin{figure}[t]
\vspace*{2mm}
\begin{center}
\includegraphics[width=8.3cm]{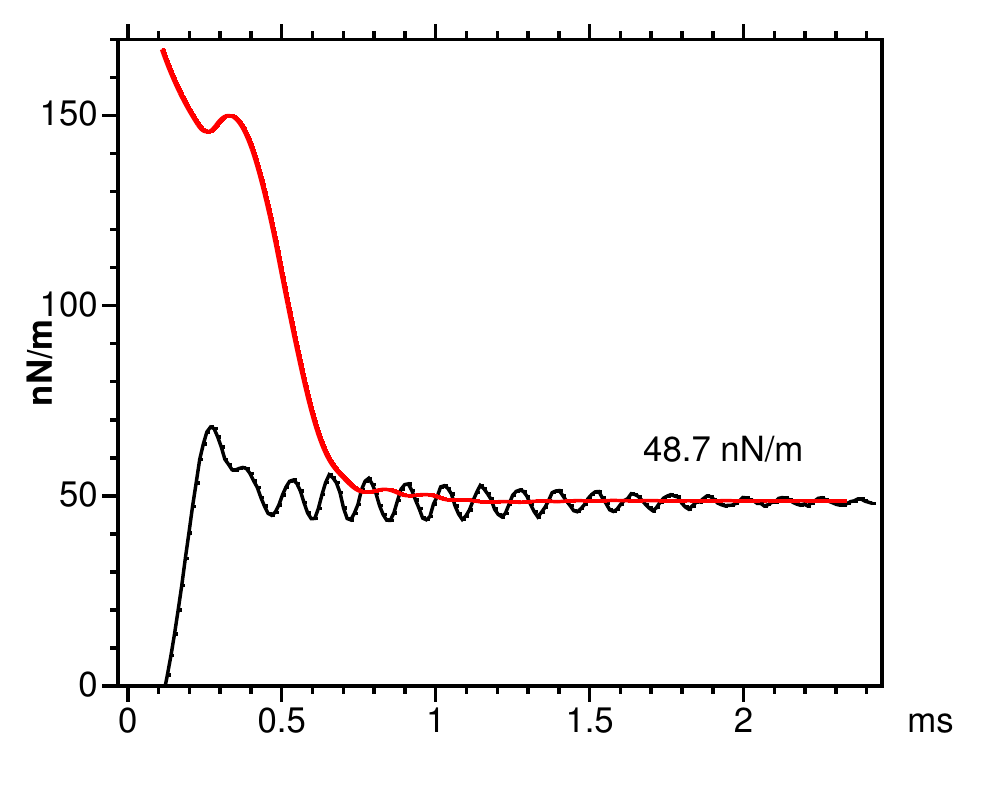}
\end{center}
\caption{Run Te: same as Fig.~\ref{fig:CurveBaseline}, but with $T_e=0.3$ eV
  instead of 0.1 eV.}
\label{fig:CurveTe}
\end{figure}

In all runs thus far presented we have assumed that the electron
temperature is 0.1 eV (the same as ion temperature). In the 700-900 km
altitude range, the electron temperature is actually often 0.2-0.3
eV. Figure \ref{fig:CurveTe} shows thrust behaviour of a run \MARKII{(named Te)} which is
otherwise identical to Baseline (V400), but the electron temperature
is set to 0.3 eV. The thrust remains essentially unchanged and the
oscillation is damped. So \MARKII{run Te suggests} that a higher electron temperature
improves stability.

\subsection{Longer runs with magnetic field}
\label{subsection:longerruns}

\begin{figure*}[t]
\vspace*{2mm}
\begin{center}
\includegraphics[width=12cm]{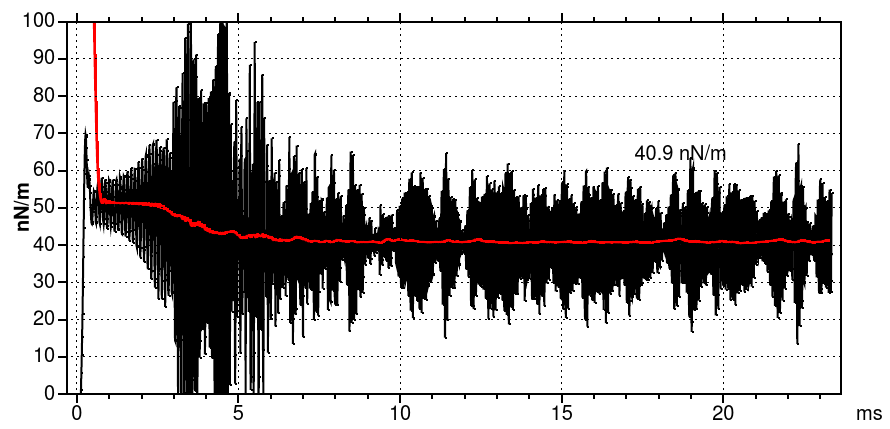}
\end{center}
\caption{Run BxLong: $B_x=30000$ nT and 23.4 ms duration.}
\label{fig:CurveBxLong}
\end{figure*}

Figure \ref{fig:CurveBxLong} shows longer time behaviour ($4\cdot
10^6$ timesteps, 23.4 ms) of the thrust in the presence of $X$-directed
magnetic field. The result indicates that after about 7 ms, the thrust
no longer decreases while intermittent unstable wave activity
continues. The asymptotic value of the thrust as determined from the
run is 40.9 nN/m which is only 11\,\% less than in V400 (Baseline) when
the difference in voltage between the runs (311 V versus 337 V) is
compensated for by a square root dependence. Hence \MARKII{the simulation suggests} that the
instability caused by an $X$-directed magnetic field decreases the
plasma brake thrust only slightly at the baseline voltage.

\begin{figure*}[t]
\vspace*{2mm}
\begin{center}
\includegraphics[width=12cm]{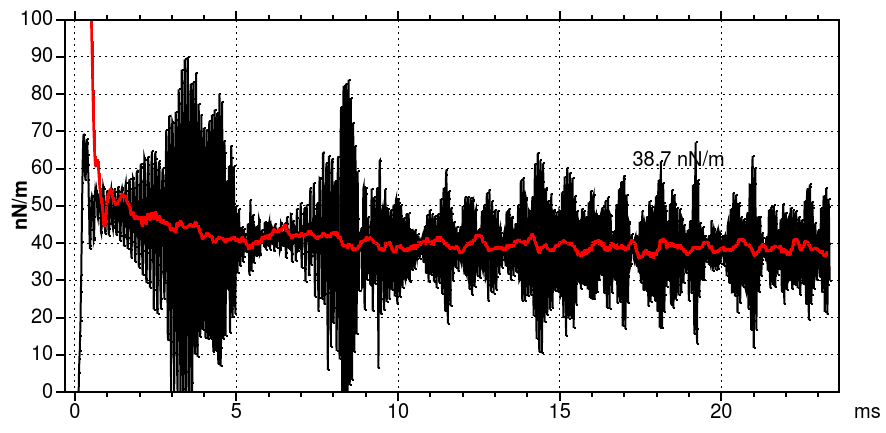}
\end{center}
\caption{Run BzLong: $B_z=30000$ nT and 23.4 ms duration.}
\label{fig:CurveBzLong}
\end{figure*}

Figure \ref{fig:CurveBzLong} shows the corresponding result for
$Z$-directed magnetic field. The behaviour is similar except that the
fluctuation spectrum extends to somewhat lower frequencies. The
final state thrust 38.7 nN/m is 17\,\% lower than voltage-corrected
Baseline.

\begin{figure*}[t]
\vspace*{2mm}
\begin{center}
\includegraphics[width=12cm]{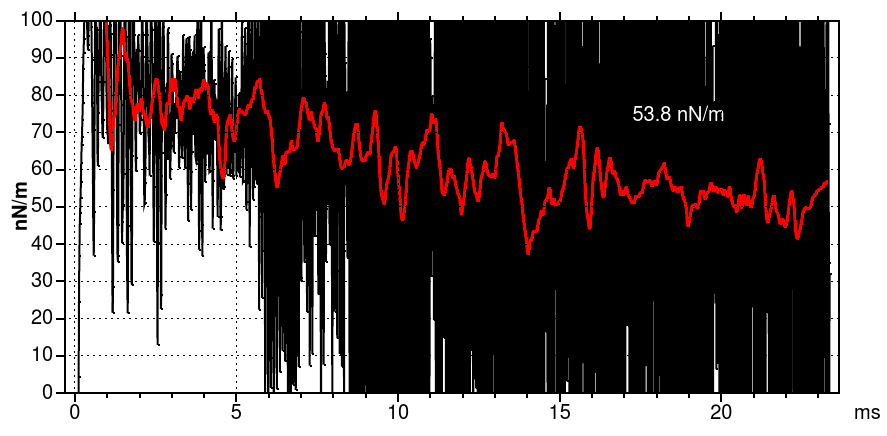}
\end{center}
\caption{Run V1000BzLong: $V_w=756.69$ V, $B_z=30000$ nT and 23.4 ms duration.}
\label{fig:CurveV1000BzLong}
\end{figure*}

Figure \ref{fig:CurveV1000BzLong} shows the result with higher voltage
(757 V which is 2.4 times larger than Baseline) and $Z$-directed
field. The fluctuations are strong and the thrust 53.8 nN/m is 27\,\%
lower than voltage-corrected Baseline. Thus the relative gap between
the Bz and ${\bf B}=0$ cases increases with voltage. The increase of
the gap is slower than linear, however, since it increases from 17\,\%
to $\sim$27\,\% if the voltage is made 2.4 times stronger. The thrust curve in
Fig.~\ref{fig:CurveV1000BzLong} might not yet be completely stabilised
at the end of the run so the thrust estimate derived from the
simulation in this case might somewhat overestimated.

\begin{figure*}[t]
\vspace*{2mm}
\begin{center}
\includegraphics[width=12cm]{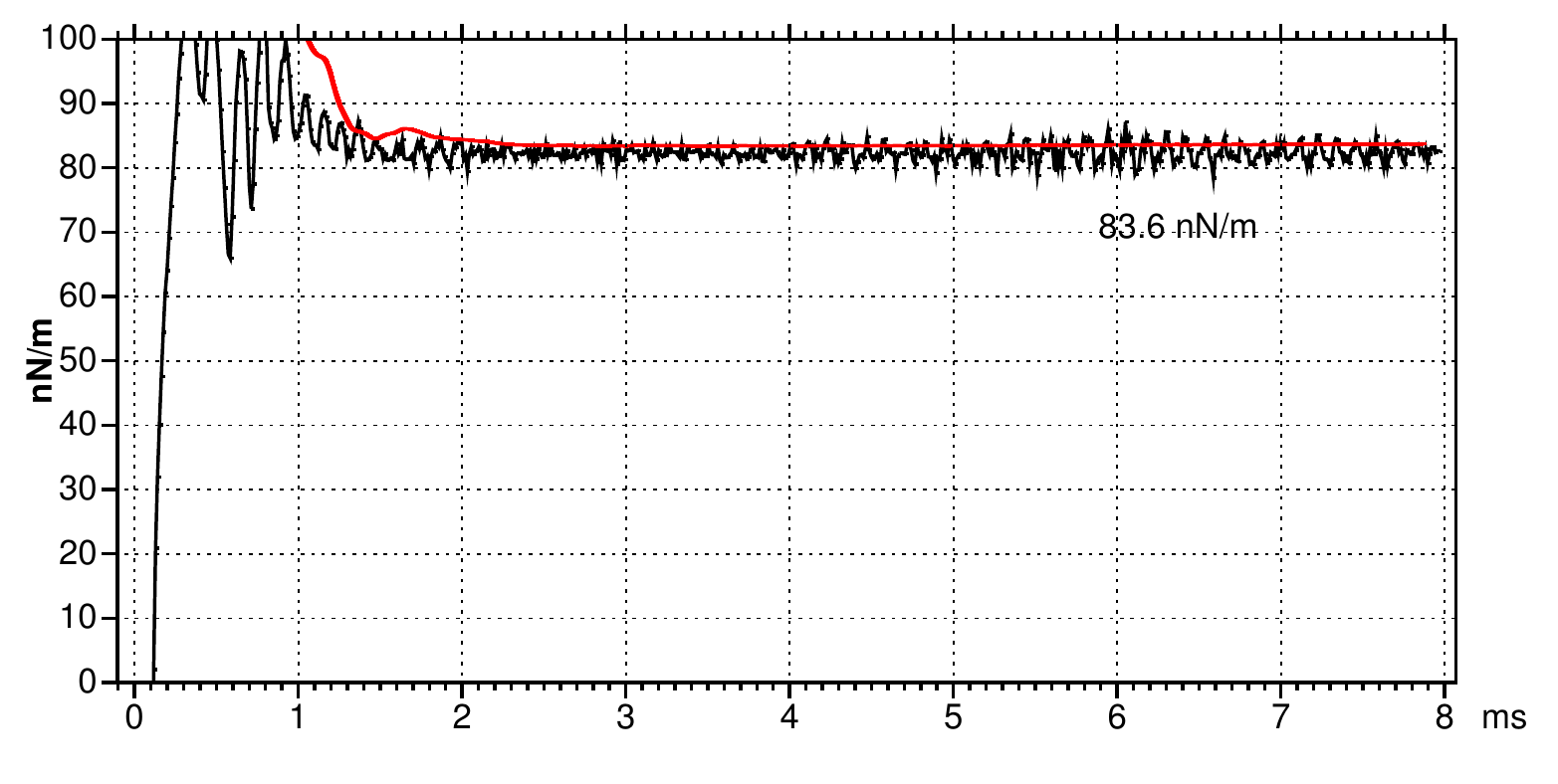}
\end{center}
\caption{Run V1000Bx: $V_w=856.07$ V, $B_x=30000$ nT and 8 ms duration.}
\label{fig:CurveV1000Bx}
\end{figure*}

The $X$-directed magnetic field case with high 856 V voltage is shown in
Fig.~\ref{fig:CurveV1000Bx}. This run provides a positive surprise
since initial transients now die away quickly after which the state is
stable and the thrust is even somewhat larger than Eq.~(\ref{eq:dFdz})
prediction. Thus, although $X$-directed field at lower voltage is
unstable and the instability lowers the thrust to a small extent (run
BxLong), at higher voltage the instability is absent and instead of a
small thrust reduction we have a minor thrust enhancement.

\begin{figure*}[t]
\vspace*{2mm}
\begin{center}
\includegraphics[width=12cm]{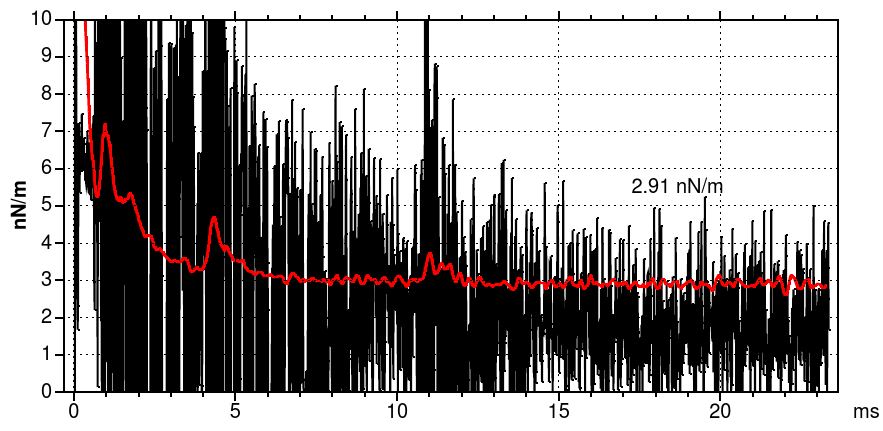}
\end{center}
\caption{Run ProtonBzLong: proton plasma, $B_z=30000$ nT and 23.4 ms duration.}
\label{fig:CurveProtonBzLong}
\end{figure*}

Next we look at the asymptotic state in the \MARKII{strongly} unstable proton case
(see subsection \ref{subsection:ionmass} above) where in addition a
destabilising Bz magnetic field exists. The result is shown in
Fig.~\ref{fig:CurveProtonBzLong}. The fluctuations are strong, but the
average magnitude of the thrust \MARKII{gets} stabilised during the
run. Thus, even \MARKII{strongly} unstable behaviour does not cause a collapse of
the ion sheath which surrounds the tether and whose spatial extent
determines the thrust. Interestingly, the determined thrust of 2.91
nN/m at $V_w=256$ V is exactly (within three decimal places) equal to
Eq.~(\ref{eq:dFdz}) prediction. Likely the ${\bf B}=0$ value of the
thrust in case of proton plasma would be slightly above
Eq.~(\ref{eq:dFdz}) prediction, while $B_z$ tends to slightly lower
the thrust.  That the effects cancel out exactly \MARKII{may be}
fortuitous.

\subsection{Miscellaneous runs}

\label{subsect:misc}
We also made a run where the magnetic field components had equal
values: $B_x=B_y=B_z=30000$ nT$/\sqrt{3}$. The result is an intermediate
case of the Bx, By and Bz runs: a very slowly growing oscillation
appears. It therefore seems that the behaviour is stable if the magnetic field
is predominantly in the $Y$ direction and unstable otherwise. The
result suggests that if the magnetic field has general orientation,
the behaviour can be qualitatively interpolated from the X, Y and Z
directed runs.

When the plasma density $\nzero$ is changed, all spatial scales in the
electrostatic PIC simulation model scale naturally by the electron
Debye length i.e.~as $\sim 1/\sqrt{\nzero}$. The only thing that
breaks this scale invariance is the fixed value of the tether's
effective electric radius $r_w^{*}$. However, the effect of $r_w^{*}$
is minor because it enters in Eq.~(\ref{eq:dFdz}) only
logarithmically. Thus we can say that changing the plasma density
$\nzero$ does not affect the stable/unstable nature of the solution
and the thrust scales nearly as proportional to $\sqrt{\nzero}$. As a
consistency check, we made one run to verify this behaviour.

We also performed a high voltage run with $Y$-directed field, V1000By
(Table \ref{tab:RunList}) with short 2.43 ms duration because only a
weak stable oscillation is present and no instability is seen. As in
the corresponding lower voltage run By, the determined thrust 80.7
nN/m is in close agreement with Eq.~\ref{eq:dFdz}.

\section{Discussion}

In Eq.~(\ref{eq:dFdz}) the thrust is nearly linearly proportional to
the ion mass and the simulations are in agreement with this at least
qualitatively. 

At $\sim 320$ V, the runs BxLong and BzLong suggest that although the
plasma sheath is unstable when an $X$ or $Z$ dominant magnetic field is
present, the thrust is reduced by the instability only modestly (11\,\%
and 17\,\%, respectively). When the voltage is increased to $\sim 800$
V, a $Z$-directed field is again unstable (V1000BzLong) and the thrust
reduction is larger, $\sim 27$\,\%. In case of $X$-directed field,
however, the instability is absent and there is actually a small
thrust enhancement (V1000Bx). A $Y$-directed field is always stable and
there is neither thrust reduction nor enhancement.

In other words, Eq.~(\ref{eq:dFdz}) is in most cases able to predict
the thrust well. The only exception is that if the magnetic field is
aligned with the tether ($Z$-directed case), then there is a moderate
thrust reduction which increases with voltage. The relative reduction
is 17\,\% at 320 V and $\sim 27$\,\% at 760 V for $Z$-directed magnetic
field.

When ${\bf B}=0$, the flow is stable with oxygen plasma and unstable
with proton plasma (runs V400 and Proton). When an
instability sets in, the ion cloud formed by ions passing near the
tether starts to oscillate at ion plasma frequency. In the oxygen case
the ion plasma oscillation is 4 times slower than in the proton
case. In the oxygen case the flow has time to move by more than one
ion sheath diameter during one ion plasma period, while in the proton
case it moves only a fraction of the sheath diameter. Thus in the
proton plasma case, if the sheath starts to oscillate, the
oscillations have more opportunities to disturb the upstream flow and
perhaps cause a positive nonlinear feedback. Maybe this \MARKII{is why}
lighter ion mass flow tends to be more unstable. Another way to arrive at a qualitatively
similar conclusion is to note that the ion bulk flow kinetic energy $e
V_i=(1/2)m_i \vzero^2$ is linearly proportional to the ion mass. Thus
the ratio $V_w/V_i$ is 16 times larger in proton plasma flow than in O$^{+}$
ion flow. Any \MARKII{voltage dependent} instability \MARKII{should
  then occur at 16 times lower voltage} in proton plasma than in O$^{+}$ plasma.

Let us consider a vertical gravity gradient stabilised plasma brake
tether using the untilted dipole approximation for Earth's
  magnetic field. The untilted dipole approximation can be used in
  this case because runs where all field components had nonzero
  values seemed to interpolate smoothly from purely aligned runs
  (Section \ref{subsect:misc}), i.e.~exact orientation of the magnetic
  field does not seem to matter. In this approximation, in equatorial orbit the magnetic field is $Y$-directed all the
time and the thrust is consequently predicted well by
Eq.~(\ref{eq:dFdz}). The same is true in polar orbit in low
latitudes. Only for the high latitude portion of a polar orbit the
thrust is somewhat smaller than what Eq.~(\ref{eq:dFdz}) predicts.

Let us look at a numeric example. At 1 kV voltage, $3\cdot 10^{10}$
m$^{-3}$ density and oxygen plasma, the predicted plasma brake thrust
from Eq.~\ref{eq:dFdz} is 85 nN/m. A 5 km long tether would produce
0.43 mN braking force which is equivalent to 13400 Ns impulse per
year. At 800 km altitude, reducing the orbital altitude by 100 km
requires 52 m/s of delta-v, thus the \MARKII{exemplary} 5 km tether (mass 0.055
kg, \citep{SeppanenEtAl2013}) could lower the orbital altitude of a
260 kg object by 100 km during one year. The gathered oxygen ion
current per tether length is given by the orbital motion limited (OML)
theory expression as
\begin{equation}
\frac{dI}{dz} = e \nzero \sqrt\frac{2e V_w}{m_i} d_w^{\rm tot}
= 8.4\cdot 10^{-8} {\rm A}/{\rm m}
\end{equation}
where $d_w^{\rm tot}$ is the total width of the four-wire tether (160
$\mu$m). Hence a 5 km tether gathers 0.4 mA current and consumes 0.4 W
power in this plasma. In reality the current is \MARKII{expected to be} somewhat
larger because of production of some secondary electrons from the
tether when hit by oxygen ions. For comparison, the maximal Lorentz
force produced by the tether \citep{CosmoAndLorenzini1997} (using the length-averaged current
  of 0.2 mA) is only 0.03 mN which is 14 times smaller
than the plasma brake force in this case.

We did not test voltages higher than about 860 V in this paper. Going
to higher voltages tends to increase the needed computing time because
the sheath becomes larger so one needs more grid cells, particles and
timesteps to model its asymptotic evolution accurately. In a practical
device, if one increases the negative voltage, at some point electron
field emission from the surface of the tether wires starts to become
an issue. Field emission adds to the ionic current gathered by the
tether and hence increases power consumption. We think that the point
where field emission starts to become an issue is larger than 1 kV but
smaller than perhaps 3-5 kV. The value also depends on the geometry
and possible coating of the tether. Before experimental knowledge
about the plasma brake is obtained, it is not necessarily well
motivated to try simulations with higher voltages than those presented
in this paper.

\subsection{Correction of an earlier result}

The reference \citet{Plasmabrake} contains an error: the quantities which
are in the present paper denoted by $\tilde{V}$ and $V_w$ were
confused with each other. Therefore the thrust versus voltage
relationship as given in \citet{Plasmabrake} was too optimistic. On
the other hand, \citet{Plasmabrake} used an earlier formula
\citep[equation 20]{paper3} whose numerical coefficient had been found
from a test particle calculation using solar wind parameters.  In the
ionospheric plasma brake case, the ratio between the tether voltage
$V_w$ and the bulk flow energy $V_i$ (in voltage units) is much larger
than in the solar wind, and consequently ions are reflected backward
more efficiently in the ionosphere than in the solar wind, as is
evident from the simulation results of the present paper. All in all,
the results of \citet{Plasmabrake} are not too far from reality: the
error resulting from confusing $\tilde{V}$ and $V_w$ and the
inaccuracy resulting from using a formula suitable for solar wind
roughly cancel out each other.

\conclusions  

According to a series of electrostatic PIC simulations performed at
different parameters, Eq.~(\ref{eq:dFdz}) can be used to predict
plasma brake thrust in ionospheric conditions. Only if the dominant
component of the magnetic field is along the tether, the thrust is
reduced, and the relative reduction grows from 17\,\% to $\sim 27$\,\%
when the voltage increases from 320 V to 760 V. The thrust reduction
is due to an instability \MARKII{which has ionic character}.

The predicted performance of the plasma brake seems promising
concerning satellite deorbiting applications. For example in O$^{+}$
plasma with $3\cdot 10^{10}$ m$^{-3}$ density and using 1 kV voltage,
a 5 km long plasma brake tether weighing 0.055 kg could produce 0.43
mN breaking force \MARKII{at altitude $\sim 800$ km,} which is enough
to reduce the orbital altitude of a 260 kg debris mass by 100 km
during one year.




\begin{acknowledgements}
The work was partly supported by Academy of Finland grant 250591.
\end{acknowledgements}



\clearpage


\clearpage


\begin{table}[t]
\caption{Simulation parameters of V400 (Baseline)run.}
\label{tab:SimParams}
\vskip4mm
\centering
\begin{tabular}{lcr}
\tophline
Parameter & Symbol& Value \\
\middlehline
Grid size    & & $512 \times 512$ \\
Grid spacing & $\Delta x$ & $7.3$ mm \\
Normalised spacing & $\Delta x/\lambda_{\rm De}$ & 0.54 \\
X grid domain & & -2.34 .. 1.4 m \\
Y grid domain & & -1.87 .. 1.87 m \\
Timestep     & $\Delta t$ & $5.84$ ns \\
Normalised timestep & $\omega_{\rm pe}\Delta t$ & 0.057 \\
Run duration & $t_{\rm max}$ & $2.43$ ms \\
Number of timesteps & & 417000 \\
Electrons per cell & $N_0$ & 500 (in plasma stream) \\
Number of particles & & $\sim\ \  52.4$ M \\
Plasma density & $\nzero$ & $3\cdot 10^{10}$ m$^{-3}$ \\
Ion mass     & $m_i$ & 16 \MARKII{amu} (O$^{+}$) \\
Plasma drift & $\vzero$ & 7.5 km/s \\
Electron temp. & $T_e$ & 0.1 eV \\
Ion temp.    & $T_i$ & 0.1 eV \\
Magnetic field & $B_x$,$B_y$,$B_z$ & 0 nT \\
Tether voltage & $V_w$ & 337.436 V \\
Tether electric radius & $r_w^{*}$ & 1 mm \\
\bottomhline
\end{tabular}
\end{table}

\begin{table*}[t]
\caption{List of performed runs. Only differences to the V400
  (Baseline) run are mentioned. Runs ending with ``g'' have larger $768\times 768$
  grid. The Long runs have 23.4 ms duration. The 'Rel.~thrust' column tells how much
  the thrust differs from Eq.~(\ref{eq:dFdz}) prediction.}
\label{tab:RunList}
\vskip4mm
\centering
\begin{tabular}{llllrr}
\tophline
Run          & Parameters     & Thrust/nNm$^{-1}$ & Eq.~(\ref{eq:dFdz})/nNm$^{-1}$ & Rel.~thrust & Nature \\
\middlehline
V50          & $V_w=41.0$ V & 7.06 & 10.6 & -33\,\% & Stable \\
V100         & $V_w=82.4$ V & 19.4 & 19.8 & -2\,\%  & Stable \\
V150         & $V_w=125$ V  & 27.3 & 26.7 & +2\,\%  & Stable \\
V200         & $V_w=168$ V  & 32.8 & 32.3 & +2\,\%  & Stable \\
V300         & $V_w=252$ V  & 40.5 & 41.2 & -2\,\%  & Steady oscillation\\
V400         & Table \ref{tab:SimParams}, $V_w=337$ V & 47.9 & 48.5 & -1\,\% & Steady oscillation\\
V500         & $V_w=424$ V  & 54.5 & 54.9 & -1\,\%  & Steady oscillation\\
V500g        & $V_w=424$ V  & 54.9 & 54.9 & 0\,\%   & Steady oscillation\\
V600g        & $V_w=512$ V  & 60.9 & 60.6 & 0\,\%   & Steady oscillation\\
V800g        & $V_w=689$ V  & 71.4 & 70.7 & +1\,\%  & Steady oscillation\\
V1000g       & $V_w=869$ V  & 80.1 & 79.5 & +1\,\%  & Steady oscillation\\
Bx           & $B_x=30 \mu$T, $V_w=333$ V & 51.1 & 48.2 & +6\,\%  & Unstable \\
BxLong       & $B_x=30 \mu$T, $V_w=311$ V & 40.9 & 46.4 & -12\,\% & Unstable \\
By           & $B_y=30 \mu$T, $V_w=338$ V & 48.5 & 48.6 & 0\,\%   & Dying oscillation\\
Bz           & $B_z=30 \mu$T, $V_w=335$ V & 47.2 & 48.4 & -2\,\%  & Unstable \\
BzLong       & $B_z=30 \mu$T, $V_w=317$ V & 38.7 & 46.9 & -17\,\% & Unstable \\
Helium       & $m_i=4$ \MARKII{amu}, $V_w=350$ V & 14.6 & 13.2 & +11\,\%     & Unstable \\
Proton       & $m_i=1$ \MARKII{amu}, $V_w=323$ V  & 5.06 & 3.24 & +56\,\%     & Unstable \\
ProtonBzLong & $m_i=1$ \MARKII{amu}, $V_w=256$ V  & 2.91 & 2.91 & 0\,\%       & Unstable \\
V1000By      & $B_y=30 \mu$T, $V=870$ V & 80.7 & 79.5 & +2\,\%    & Steady oscillation\\
V1000BzLong  & $B_z=30 \mu$T, $V_w=757$ V & 53.8 & 74.1 & -27\,\% & Unstable \\
V1000Bx      & $B_x=30 \mu$T, $V_w=856$ V & 83.6 & 78.0 & +7\,\%  & Stable \\
Te           & $T_e=0.3$ eV, $V_w=337$ V & 48.7 & 48.5 & 0\,\%    & Dying oscillation\\
\bottomhline
\end{tabular}
\end{table*}


%
%





\end{document}